\documentclass[preprint,12pt,numbers]{elsarticle}
\usepackage{graphics}
\usepackage{natbib}
\usepackage{amssymb}
\journal{Solid State Communications}
\begin{document}
\begin{frontmatter}
\title{Effect of p-d hybridization and structural distortion on the electronic properties of $AgAlM_2 (M = S, Se,Te)$ chalcopyrite semiconductors}
\author[]{S.Mishra}
\author[]{B.Ganguli\corref{cor1}}
\ead {biplabg@nitrkl.ac.in}
\cortext[cor1]{corresponding author. Tel.: +91661 2462725; fax: +91661 2462999 }
\address {National Institute of Technology, Rourkela-769008, Odisha, India}
\begin{abstract}
We have carried out ab-initio calculation and study of structural and electronic properties of $AgAlM_2$ (M = S, Se, Te) chalcopyrite semiconductors using Density Functional Theory (DFT) based self consistent Tight binding Linear Muffin Tin orbital (TB-LMTO) method. Calculated equlibrium values of lattice constants, anion displacement parameter (u), tetragonal distortion ($\eta$ = c/2a) and bond lengths have good agreement with experimental values. Our study suggests these semiconductors to be direct band gap semiconductors with band gaps  1.98 eV, 1.59 eV and 1.36 eV respectively. These are in good agreement with experimental value within the limitation of local density approximation (LDA). Our explicit study of the effects of anion displacement and p-d hybridization show that band gap increases by 9.8\%, 8.2\% and 5.1\% respectively for $AgAlM_2$ (M = S, Se, Te) due to former effect and decreases by 51\%, 47\% and 42\% respectively due to later effect.
\end{abstract}
\end{frontmatter}
\section{Introduction}
The ternary semiconducting compounds with the formula $A^IB^{III}C_2^{VI}$ and $A^{II}B^{IV}C_2^V$ such as $CuAlS_2$, $AgAlSe_2$, $CdSiP_2$, $ZnSnAs_2$ etc have been widely investigated because of their potential applications in electro-optic, optoelectronic and nonlinear optical devices. These compounds are promising candidates for solar cells \cite{1}, photovoltaic detectors, light emitting diodes \cite{2}, modulators, filters such as optical light eleminator filters \cite{3} and optical frequency conversion applications in solid state based tunable laser systems \cite{4}. Among the two chalcopyrite type crystals the most intensive studies have been carried out in the series of $I-III-VI_2$ family compounds where Cu is involved as a group I element \cite{5,6}. Very few works have appeared in literature where the group I element is Ag \cite{7,8}. No detailed experimental and theoretical study of  $A^IB^{III}C_2^{VI}$ type semiconductors where group I element is Ag and group III element is Al have been carried out. The substitution of a group I and a group III elements give an average result similar to that of a group II element in II-VI compounds. So in case of $AgAlM_2$ the presence of three different kind of atoms provide desired structural, electronic and optical properties to satisfy the criteria for suitable electronic devices \cite{3,9}.\\
$I-III-VI_2$ type chalcopyrite semiconductors are also important from chemistry and physics point of view in many aspects. Due to presence of group IB transition metal the d-electrons participate in chemical bonding. Therefore it is important to know how much contribution of d-electrons are there and how they influence the electronic properties. Many defect systems of this type of chalcopyrites have also been synthesized and studied \cite{10,11,12}. Due to defect structure, the system is more porous and so they have rich chemistry. Therefore the defect systems can be doped by suitable elements to synthesize new materials \cite{13,14}.\\
Jaffe and Zunger \cite{15} have proposed theory for band gap anomalies for  $A^IB^{III}C_2^{VI}$ chalcopyrite semiconductors relative to the binary analogous with the detailed study of Cu-based chalcopyrites. They have shown significant effects of structural distortion on the band gap. Their calculation was based on the Potential-Variational-Mixed Basis (PVMB) taking the experimental lattice parameters. Similar effects are also expected for Ag-based chalcopyrite but there are no systematic study carried out for these compounds. We fill up this gap. We have carried out a detailed band structure calculation using localized Tight Binding Linear Muffin-Tin Orbital (TB-LMTO) method \cite{16,17}. The ab-initio calculation is based on Density Functional Theory (DFT) \cite{18,19}. Unlike taking experimental lattice constants we have calculated total energy to find the equillibrium lattice constants, tetragonal distortion ($\eta$) and anion displacement parameter (u). \\
Shay and Kasper \cite{20} have shown that for Cu-based chalcopyrite the reduction in the band gaps relative to their binary analogous is co-related with the existance of d-bonding. Their calculation was based on the study of spin-orbit splitting. This is further varified by Jaffe and Zunger \cite{15}. Artus and his group \cite{8} have measured spin-orbit interaction at the fundamental gap for $AgGaSe_2$. They found that this splitting is smaller than the splitting for analogous ideal Zinc blende structure. They argued that this is because of p-d hybridization. We have also shown with detailed calculation of partial density of states (PDOS) that hybridization of d-orbital of Ag with p-orbital of anion have significant contribution to band gap. In TB-LMTO method, the basis functions are localized. Therefore, very few basis functions are required to represent the highly localized d-orbital of Ag in the systems under study. Hence the calculation is not only cost effective, it gives also the accurate result.  
\section{Methodology}
To start with, for our electronic structure calculations we have used the well established TB-LMTO method, discussed in detail elsewhere \cite{16,17}. Electron correlations are taken  within LDA of DFT \cite{18,19}. We have used the von Barth-Hedin exchange \cite{21} with 512 {\bf k}-points in the irreducible part of the Brillouin zone. The basis of the TB-LMTO starts from the minimal set of muffin-tin orbitals of a KKR formalism and then linearizes it by expanding around a `nodal' energy point $E_{\nu \ell}^\alpha$. The wave-function is then expanded in this basis :
\begin{eqnarray}
\Phi_{j {\bf k}}(\bf r) = \sum_L \sum_\alpha c_{L\alpha}^{j {\bf k}} \left[ \phi_{\nu L}^{\alpha}(\bf r) +\sum_{L'} \sum_{\alpha'} h_{LL'}^{\alpha \alpha'}(\bf k)\dot{\phi}_{\nu L'}^{\alpha '}(\bf r)\right]
\end{eqnarray}
where,
\begin{eqnarray*}
\phi_{\nu L}^ \alpha({\mathbf{r}}) & = & \imath^{\ell}\ Y_L(\hat{r})\ \phi_\ell^\alpha(r,E_{\nu\ell}^\alpha) \\
{\mathaccent 95 \phi}_{\nu L}^\alpha(\mathbf{r}) & = & \rule{0mm}{6mm}\imath^{\ell}\ Y_L(\hat{r})\ \frac{\partial
\phi_\ell^\alpha(r,E_{\nu \ell}^\alpha)}{\partial E} \\
h_{LL'}^{\alpha\alpha '}(\mathbf{k}) & = & (C_L^\alpha - E_{\nu \ell}^\alpha)\ \delta_{LL'}\delta_{\alpha\alpha'} 
 + \sqrt{\Delta_{L}^\alpha}\ S_{LL'}^{\alpha\alpha'}(\mathbf{k})\ \sqrt{\Delta_{L'}^{\alpha'}} \\
\end{eqnarray*}
\noindent $C_L^\alpha$ and $\Delta_L^\alpha$ are TB-LMTO potential parameters and  $S_{LL'}^{\alpha\alpha'}(\mathbf{k})$ is the structure matrix.
\section{Result and discussion}
\subsection{Structural properties :}
\label{3.1}
\label{Structural properties}
The tetragonal unit cell of a typical chalcopyrite semiconductor consists of two Zinc blende unit cells.
\begin{figure}[h]
\centering
{\resizebox{3.0cm}{3.0cm}{\includegraphics{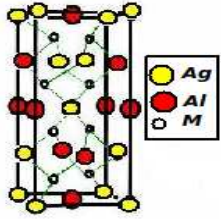}}}
\caption{One unit cell of the chalcopyrite lattice}
\end{figure}
There are four Ag atoms, four Al atoms and eight M (M = S, Se, Te) atoms per unit cell. The positions of the various atoms in the tetragonal unit cell are : A(Ag) :\ 0.0\ 0.0\ 0.0; B(Al):\ 0.0\ 0.0\ 0.5; C(M): u\ 0.25\ 0.125, where `u' is anion displacement parameter. In an ideal zinc-blende structure of binary compound like GaAs each anion has four similar cations as nearest neighbour. So the bond lengths are equal ($R_{AC} =  R_{BC}$) and the charge distribution is identical around each bond. Therefore in binary compound, having zinc blende structure, u is 0.25 and $\eta$ = 1. We refer this as ideal case. But in chalcopyrite system under study each anion has two Ag and two Al cations as nearest neighbours as shown in figure 1. Due to dissimilar atoms as neighbours the anion acquires an equilibrium position closer to one pair of cation than to other. This new position of anion is called anion displacement  which is the main cause for bond alternation ($R_{AC} \neq R_{BC}$). We refer this case as `non-ideal' case ($u \neq 0.25$ \& $\eta \neq 1$). The quantity $u - \frac{1}{4} = (R^2_{AC} - R^2_{BC})/a^2$ measures the extent to which the bond will alter in a system. The bond alternation gives rise to structure anomalies relative to the ideal zinc blend structure and it has significant effect on band gap. \\
 For self consistent calculation we introduce empty spheres because the packing fraction is low due to tetrahedral co-ordination of ions. We ensure proper overlap of muffin tin spheres for self consistency and the percentage of overlap is found.
\begin{table}
\caption{\label{label} Structural parameters including bond lengths of $AgAlM_2$. $a_{exp}$, $c_{exp}$, $u_{exp}$ are experimental parameters \cite{22} and $u_{other}$: other calculated parameters \cite{15}.}
\begin{tabular}{@{}lllllllllll}
\hline
Systems&$a$&$c$&$a_{exp}$&$c_{exp}$&$\eta$&$u$&$u_{exp}$&$u_{other}$&$R_{AC}$&$R_{BC}$ \\
&$(\AA{})$&$(\AA{})$&$(\AA{})$&$(\AA{})$& & & & &$(\AA{})$&$(\AA{})$\\
\hline
$AgAlS_2$&5.48&10.90&5.72$^{a}$&10.13$^{a}$&0.994&0.265&0.290$^{a}$&0.288$^{b}$&2.42&2.32\\
$AgAlSe_2$&5.78&11.52&5.95$^{a}$&10.75$^{a}$&0.996&0.263&0.270$^{a}$&0.287$^{b}$&2.54&2.45\\
$AgAlTe_4$&6.22&12.28&6.29$^{a}$&11.83$^{a}$&0.987&0.261&0.260$^{a}$&0.285$^{b}$&2.72&2.64\\
\hline
\end{tabular}\\
$^{a}$ Ref.\cite{22}\\
$^{b}$ Ref.\cite{15}
\end{table}
Table 1 shows the structural parameters including bond lengths of the systems under study. Our obtained lattice parameters have good agreement with experimental results \cite{22} as well as other computational work \cite{15}. Jaffe and Zunger \cite{15} have calculated `u' and bond lengths using experimental values of the lattice parameters obtained by Hahn \cite{22} and conservation of tetrahedral bonds (CTB) plus $\eta = \eta_{tet}$ rule. In our work, we have calculated the ground state equilibrium value of lattice parameters, u, $\eta$ and bond lengths by minimizing the total energy using first principle procedure. Our result shows that u increases from the ideal value $u = 0.25$ for all the three chalcopyrites which is in agreement with the other work of Jaffe and Zunger \cite{15}. Therefore, we find $R_{Ag-M} > R_{Al-M}$. The x-coordinate of M-atoms is u which is a function of lattice constants a and c. Thus the two near-neighbour distances A-C and B-C are given by equation 2 and equation 3.\\ 
\begin{eqnarray}
R_{AC} = a\left[u^2 +\frac{ (1+\eta^2)}{16}\right]^{\frac{1}{2}}\\
R_{BC} = a\left[\left(u-\frac{1}{2}\right)^2 +\frac{ (1+\eta^2)}{16}\right]^{\frac{1}{2}} 
\end{eqnarray}
\subsection{Electronic properties}
\label{3.2}
(i) $AgAlS_2$ : The band structure and total density of states (TDOS) (figure 2) show four major subbands of different band widths below fermi level $E_F$ within the energy range 0 to -13.84 eV. The first three subbands having band widths 2.29 eV, 0.91 eV, 3.17 eV respectively below the valence band maxima are separated by very narrow band gaps of 0.22 eV and 0.11 eV.
\begin{figure}
  \begin{center}
  \setlength{\tabcolsep}{-1.70cm}
    \begin{tabular}{cc}
      \rotatebox{-90} {\resizebox{60mm}{!}{\includegraphics{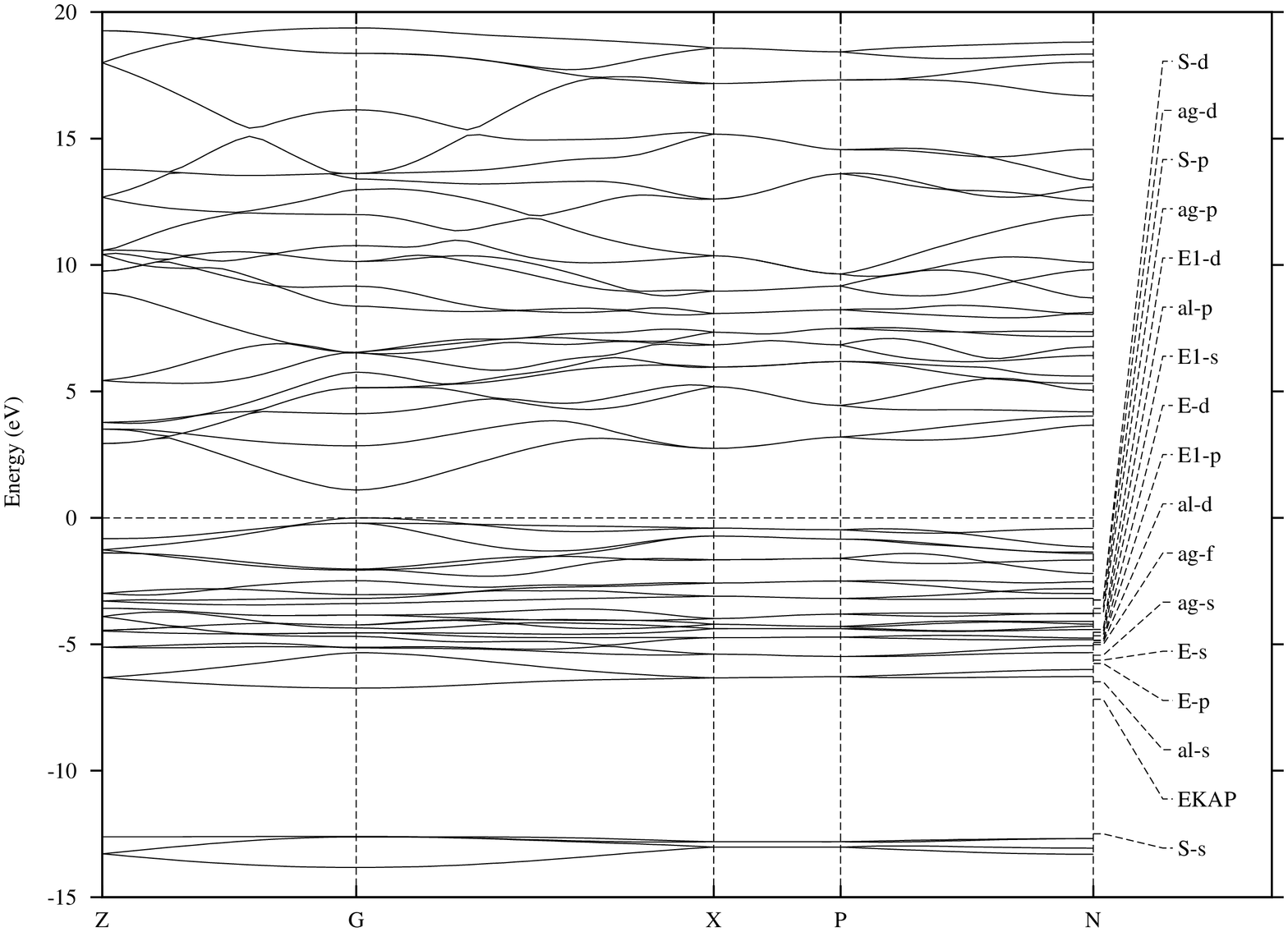}}} &
        \raisebox{1.3mm}[0pt]{\rotatebox{-90}{\resizebox{61.9mm}{!}{{\includegraphics{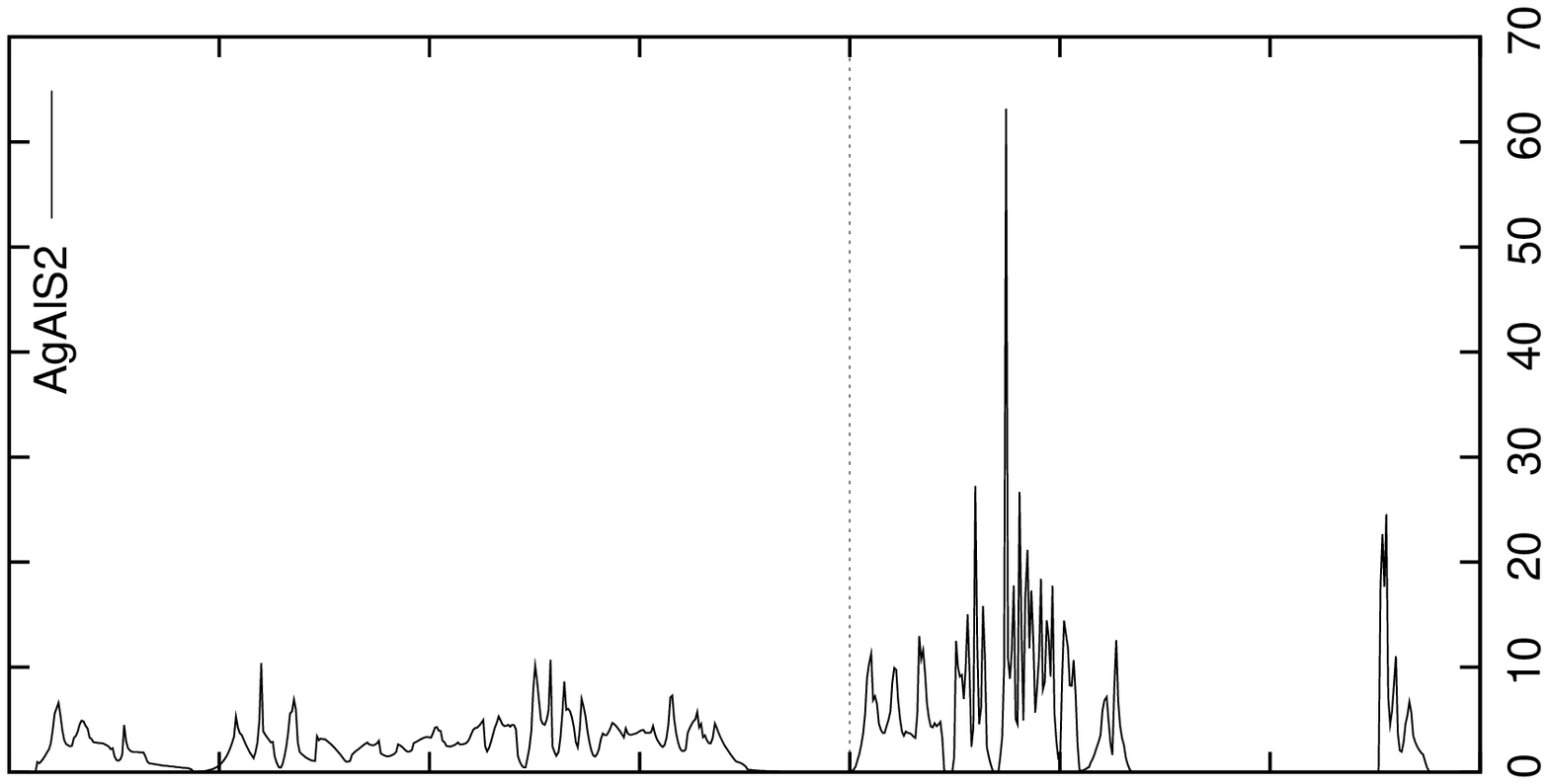}}}}} \\
        \end{tabular}
    \caption{Band structure and TDOS for non-ideal AgAlS2}
    \label{test4}
  \end{center}
\end{figure}   
\begin{figure}[h]
\centering
{\rotatebox{-90}{\resizebox{5.0cm}{10.0cm}{\includegraphics{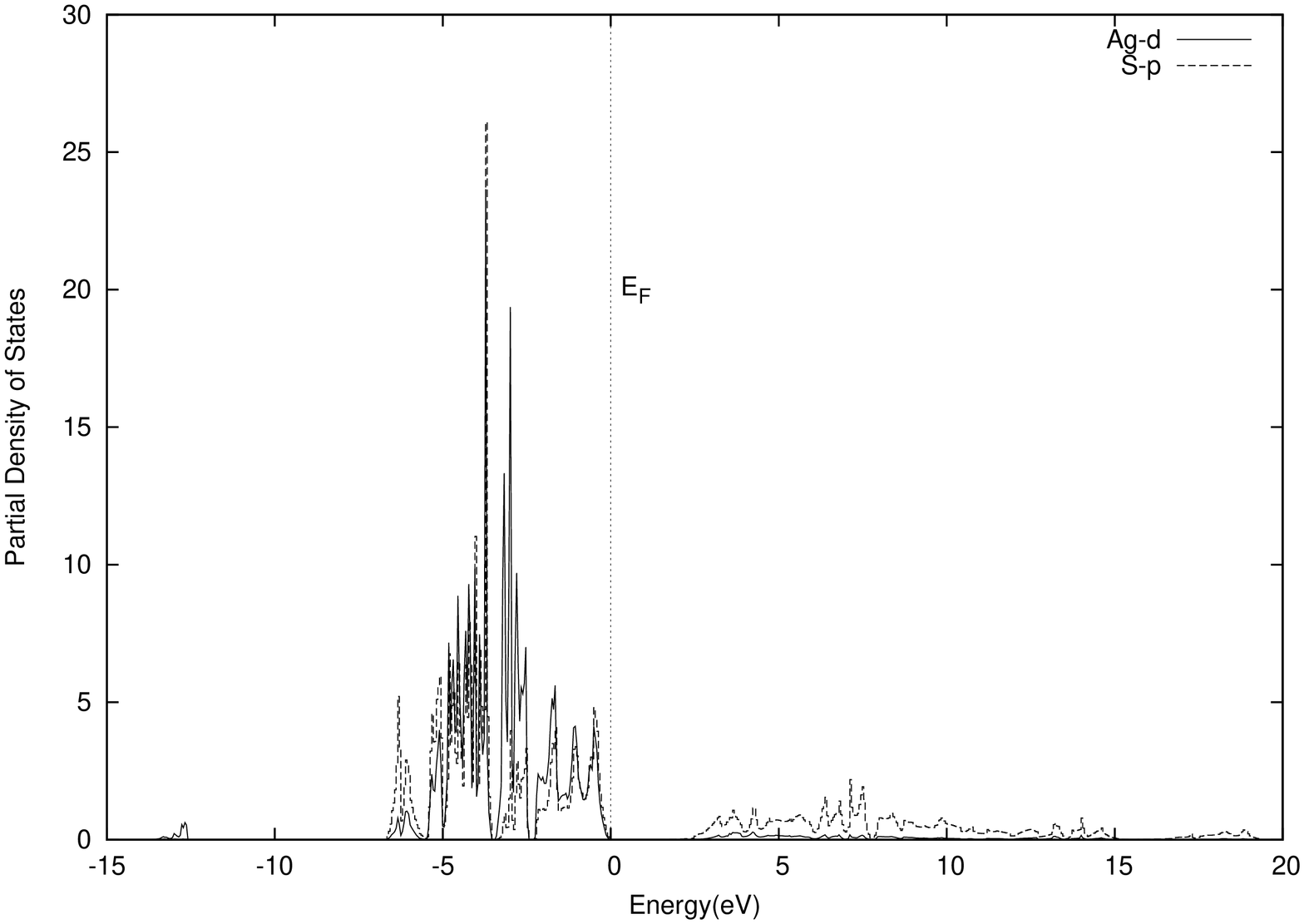}}}}
\caption{Partial density of states (PDOS) of Ag-d and S-p for non-ideal $AgAlS_2$ }
\end{figure}
The lowest subband having band width 1.24 eV is formed mainly due to the contribution of Sulfur 3s states. There is a large band gap of $\simeq$ 5.90 eV between the lowest and third subband. Partial density of states (PDOS) for Ag-d and S-p are plotted in figure 3 to study the orbital character and nature of hybridization. These figures show that Ag-d and S-p hybridization contribute to upper valence band near fermi level. There is a very weak contribution of Ag-d states in conduction band. Calculated PDOS for s,p,d states of Ag, Al, S show that the  contribution to conduction band is due to Ag-p, Al-p, S-d and S-p states.\\     
(ii) $AgAlSe_2$ : Figure 4 for band structure and TDOS show four subbands of different band widths below fermi level $E_F$ within the energy range 0 to -13.73eV.
\begin{figure}
  \begin{center}
  \setlength{\tabcolsep}{-1.70cm}
    \begin{tabular}{cc}
      \rotatebox{-90} {\resizebox{60mm}{!}{\includegraphics{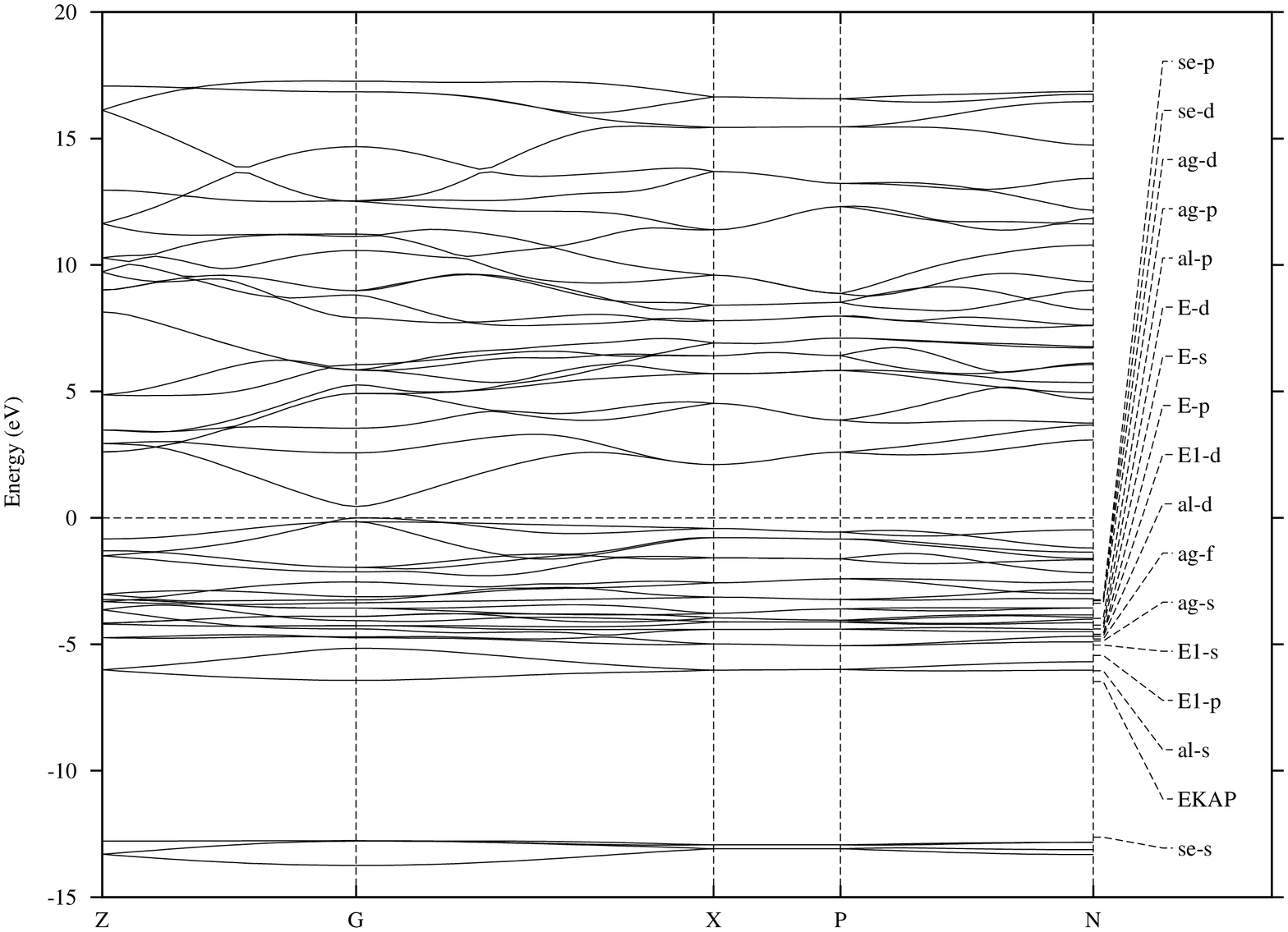}}} &
        \raisebox{1.3mm}[0pt]{\rotatebox{-90}{\resizebox{61.9mm}{!}{{\includegraphics{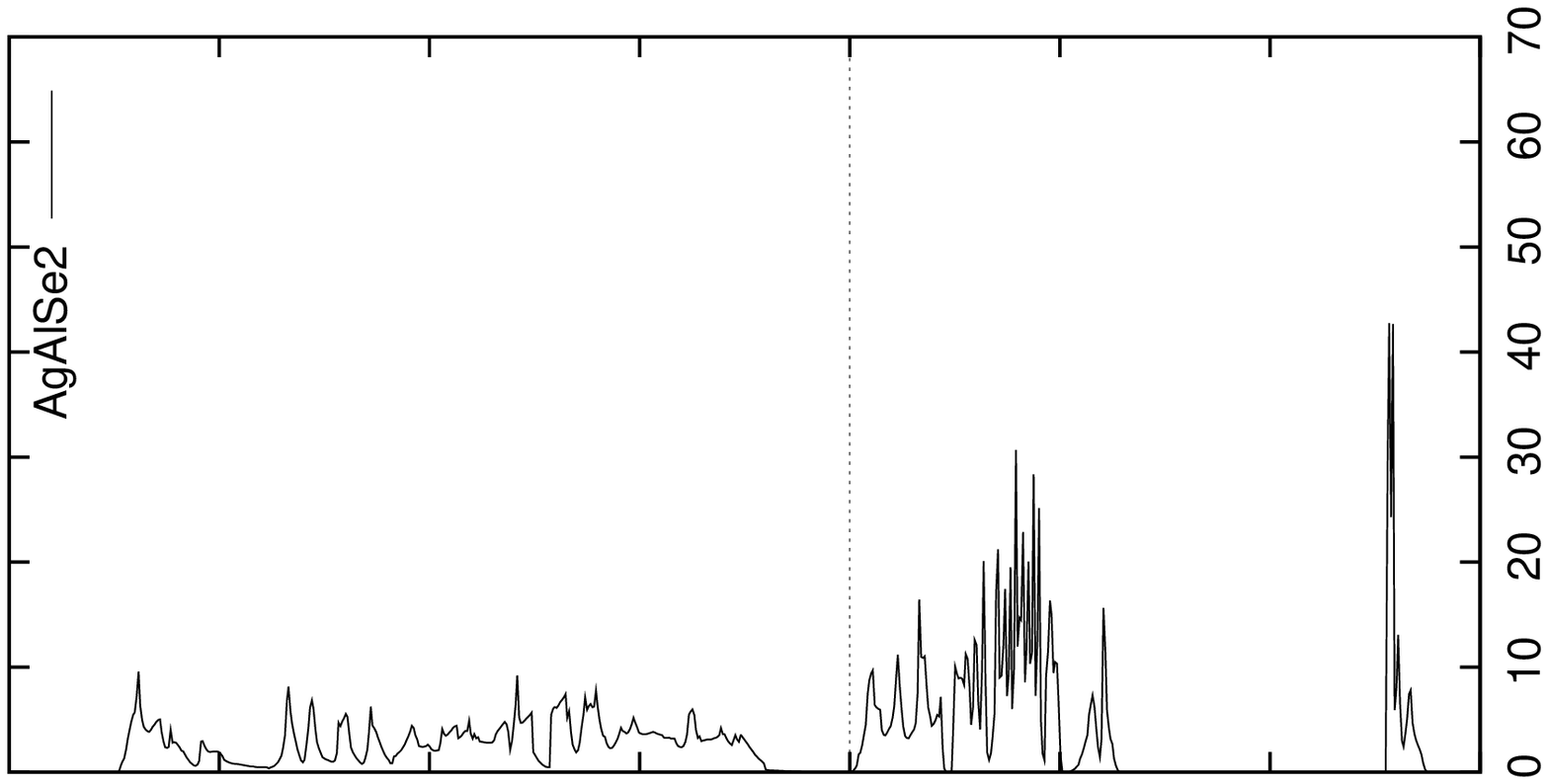}}}}} \\
        \end{tabular}
    \caption{Band structure and TDOS for non-ideal AgAlSe2.}
    \label{test4}
  \end{center}
\end{figure}   
The lowest subband is formed due to Se-4s states within the energy ranges -12.71 to -13.73 eV. There is a large band gap of nearly 6.24 eV between the lowest and just above subband. The second lowest band is mainly formed due to the contribution of Al-s and Se-p states.
\begin{figure}[h]
\centering
{\rotatebox{-90}{\resizebox{5.0cm}{10.0cm}{\includegraphics{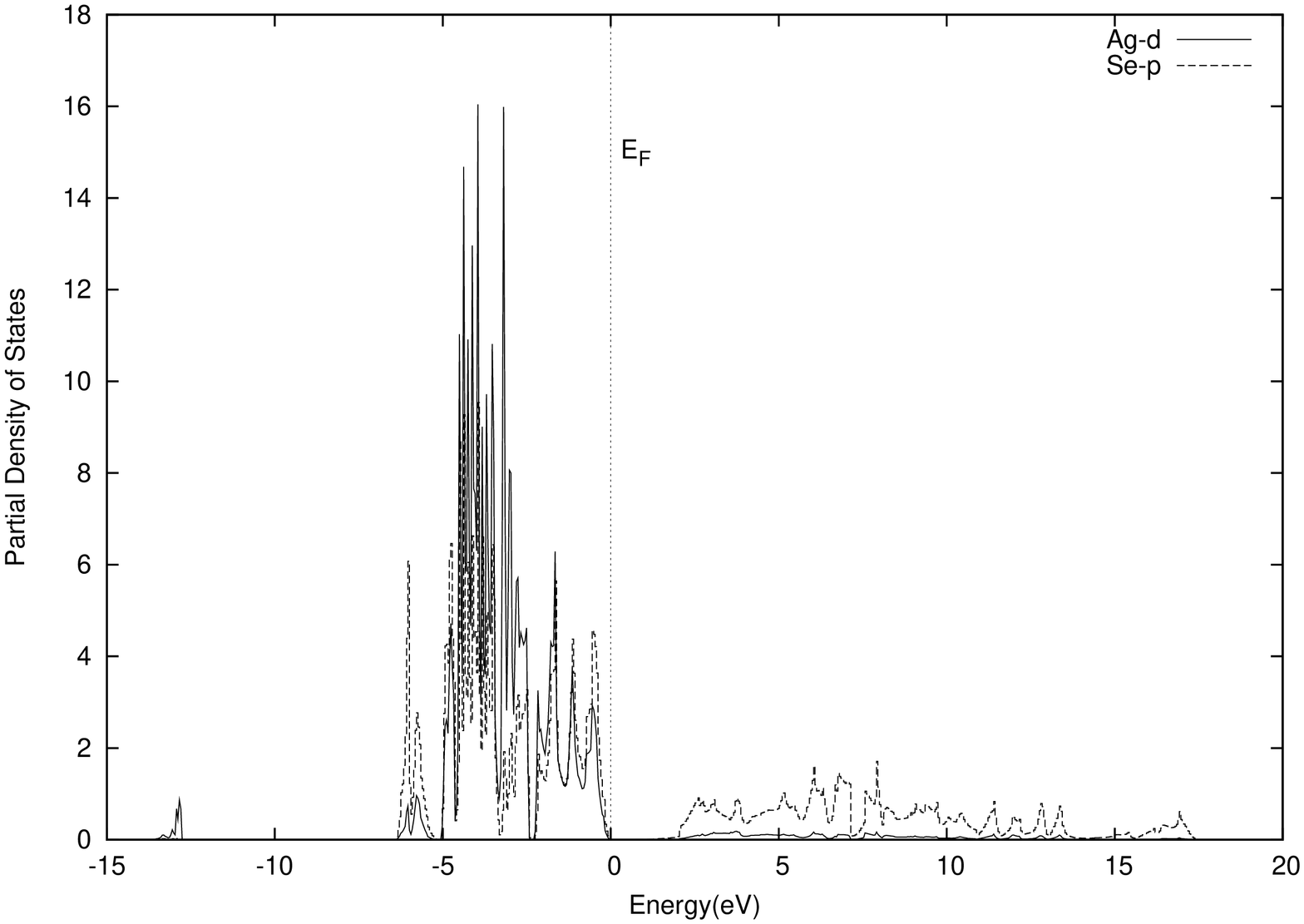}}}}
\caption{PDOS of Ag-d and Se-p for non-ideal $AgAlSe_2$}
\end{figure}
A very narrow band gap of about 0.36 eV separates the second lowest subband from the third one. The main contribution to the third lowest subband is due to Al-p, Se-p and Ag-d states. The upper most valence band is separated by the previous subband by a band gap of about 0.24 eV. Figure 5 for PDOS of Ag-d and Se-p depicts that all the Ag character is concentrated at the upper valence band leaving a very negligible amount in conduction band. The upper valence band is dominated by Ag-d and Se-p hybrid character. Calculated PDOS for s,p,d states of Ag, Al, Se show that the conduction band has main contribution from Ag-p, Al-p and Se-p and a very less contribution from Ag-s and Al-s states.\\
(iii) $AgAlTe_2$ : Unlike  $AgAlS_2$ and $AgAlSe_2$, figure 6 for electronic band structure and TDOS of $AgAlTe_2$ shows three major subbands of different band widths below fermi level $E_F$ within the energy range 0 to -12.03 eV.
\begin{figure}
  \begin{center}
  \setlength{\tabcolsep}{-1.70cm}
    \begin{tabular}{cc}
      \rotatebox{-90} {\resizebox{60mm}{!}{\includegraphics{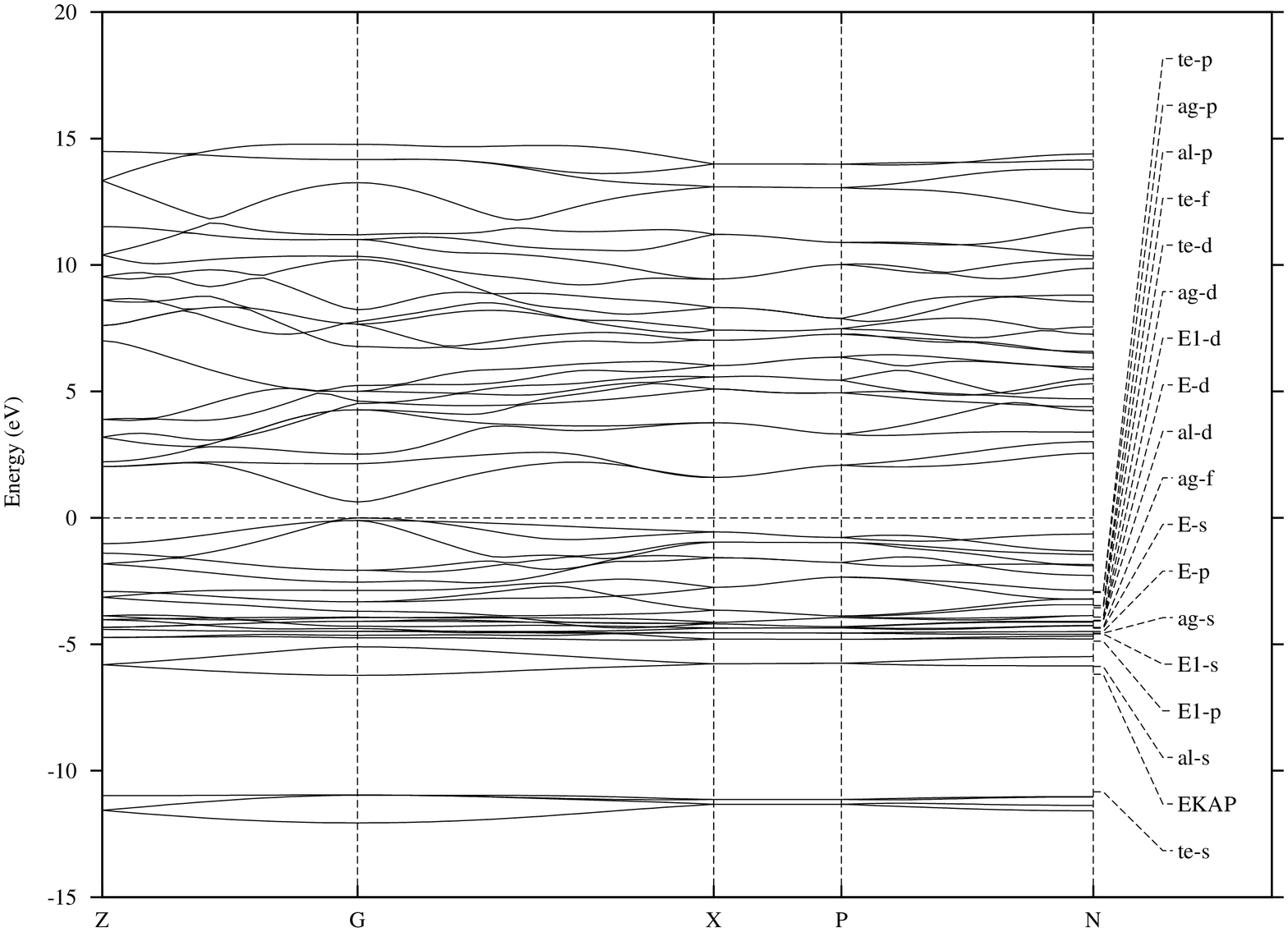}}} &
        \raisebox{1.3mm}[0pt]{\rotatebox{-90}{\resizebox{61.9mm}{!}{{\includegraphics{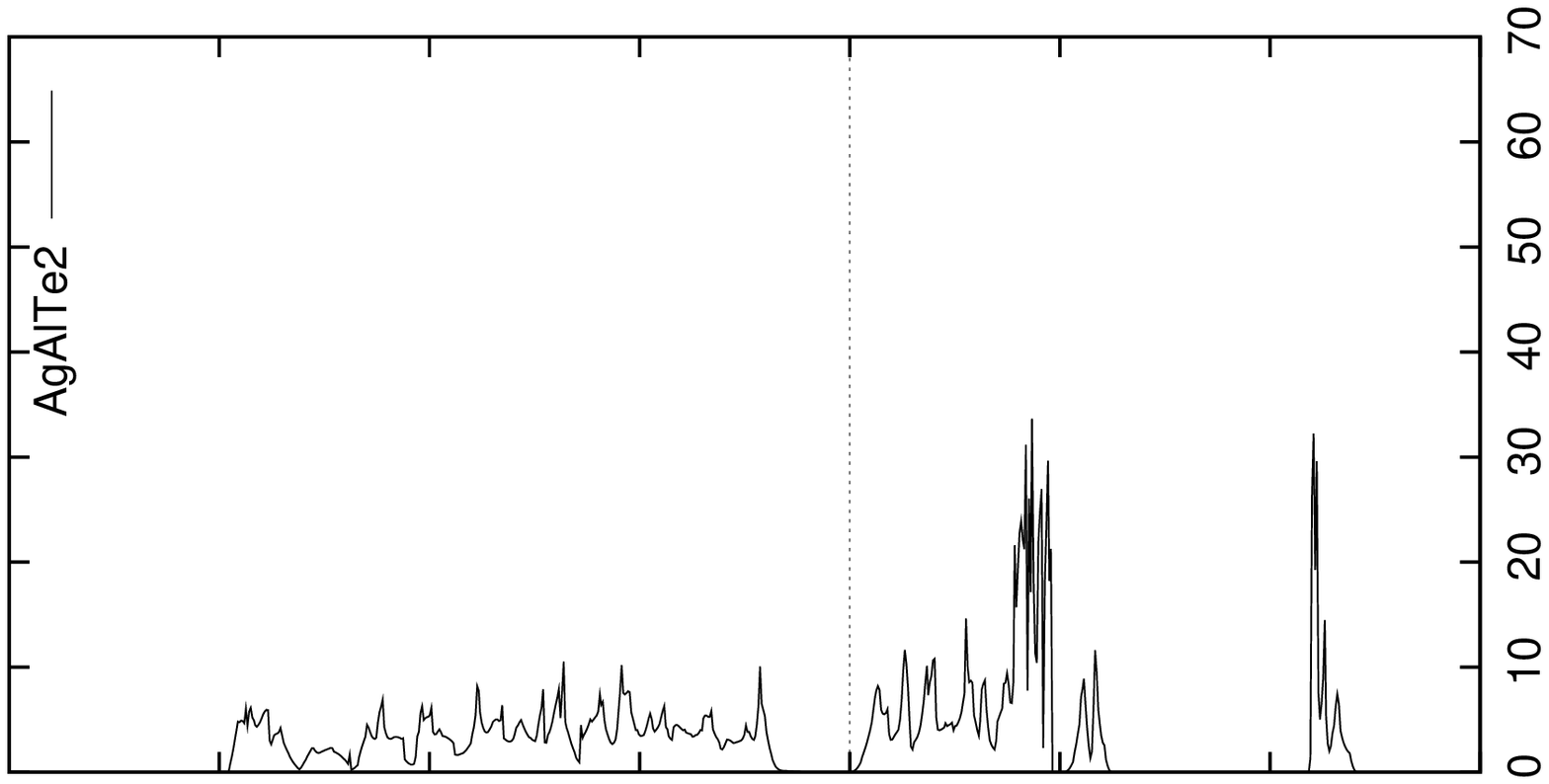}}}}} \\
        \end{tabular}
    \caption{Band structure and TDOS for non-ideal AgAlTe2 .}
    \label{test4}
  \end{center}
\end{figure}   
The upper most valence band is in the region of 0 to -4.63 eV. It mainly consists of Ag-d with some contribution from Te-p states. There is a very narrow band gap of about 0.48 eV between the upper most valence band and the previous subband.
\begin{figure}[h]
\centering
{\rotatebox{-90}{\resizebox{5.0cm}{10.0cm}{\includegraphics{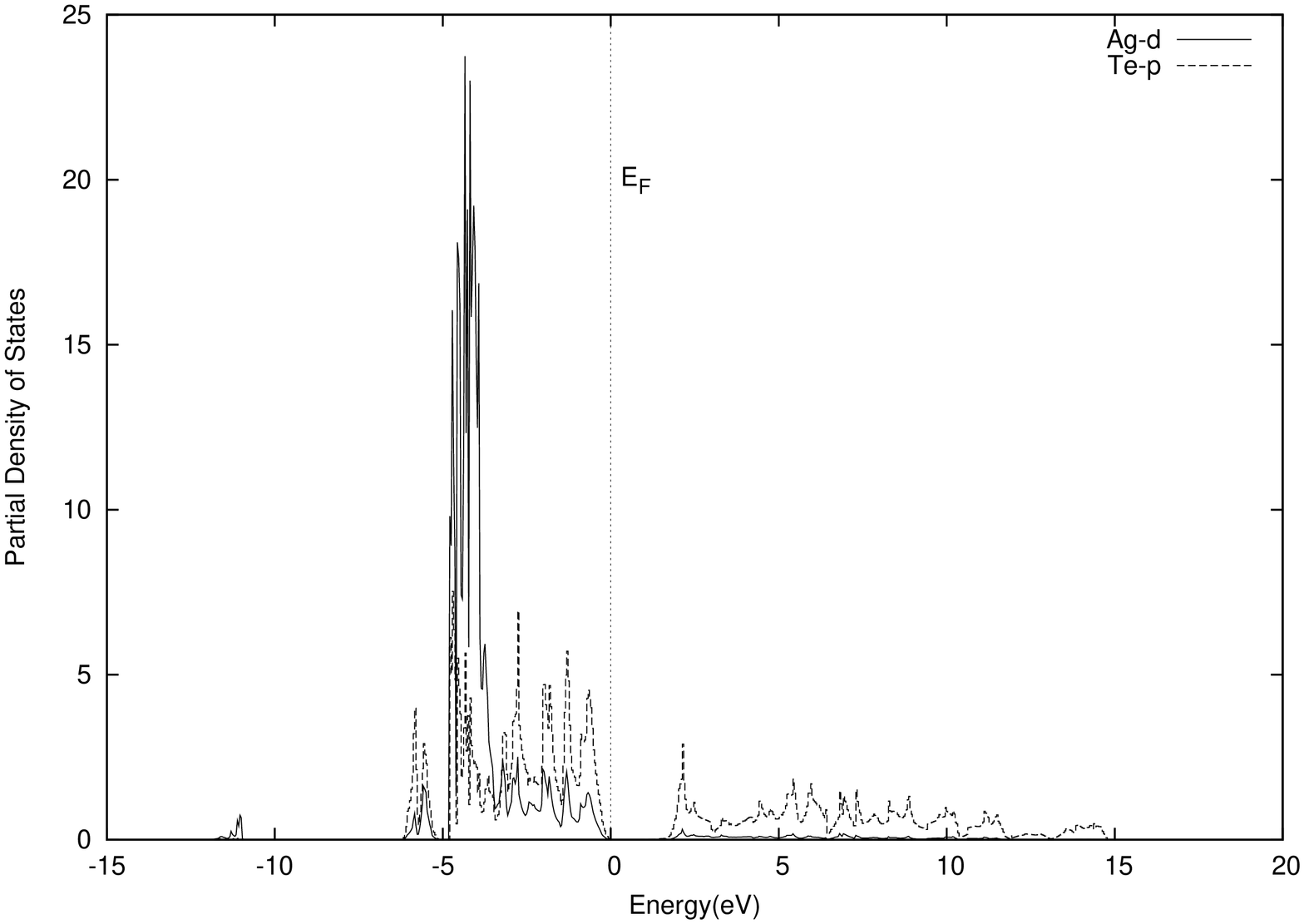}}}}
\caption{PDOS of Ag-d and Te-p for non-ideal $AgAlTe_2$ }
\end{figure}
This second subband is in the region -5.11 to -6.24 eV. It is mainly formed due to the contribution of Al-s and less contribution from Te-p states. Whereas the lowest or third subband is formed in the region of -11.0 to -12.03 eV due to the contribution of Te-5s states. Figure 7 for PDOS shows Ag-d character is concentrated in the upper most valence band and a very negligible amount of Ag-d character is in conduction band. Like in case of $AgAlSe_2$ the main contribution to upper most valence band is due to Ag-d and Te-p hybrid orbitals. Calculated PDOS for s,p,d states of Ag, Al, Te depict that the main contribution to conduction band is from Te-p, Ag-s/p and Al-s/p states.\\
In all the cases valence band maximum (VBM) and conduction band minimum (CBM) are located at center of Brillouin zone denoted as `G'($\Gamma$ point). This  indicates that they are all direct band gap compounds having band gaps 1.98 eV, 1.59 eV and 1.36 eV respectively. It is known that LDA underestimates band gap by 30\%. If we correct this error, our results are in good agreement with experimental results i.e. 3.13 eV, 2.55 eV and 2.27 eV respectively for $AgAlS_2$, $AgAlSe_2$ and $AgAlTe_2$ \cite{23,24}. 
\subsection{Effect of p-d hybridization on electronic properties}
\begin{figure}
  \begin{center}
  \setlength{\tabcolsep}{-1.70cm}
    \begin{tabular}{cc}
      \rotatebox{-90} {\resizebox{60mm}{!}{\includegraphics{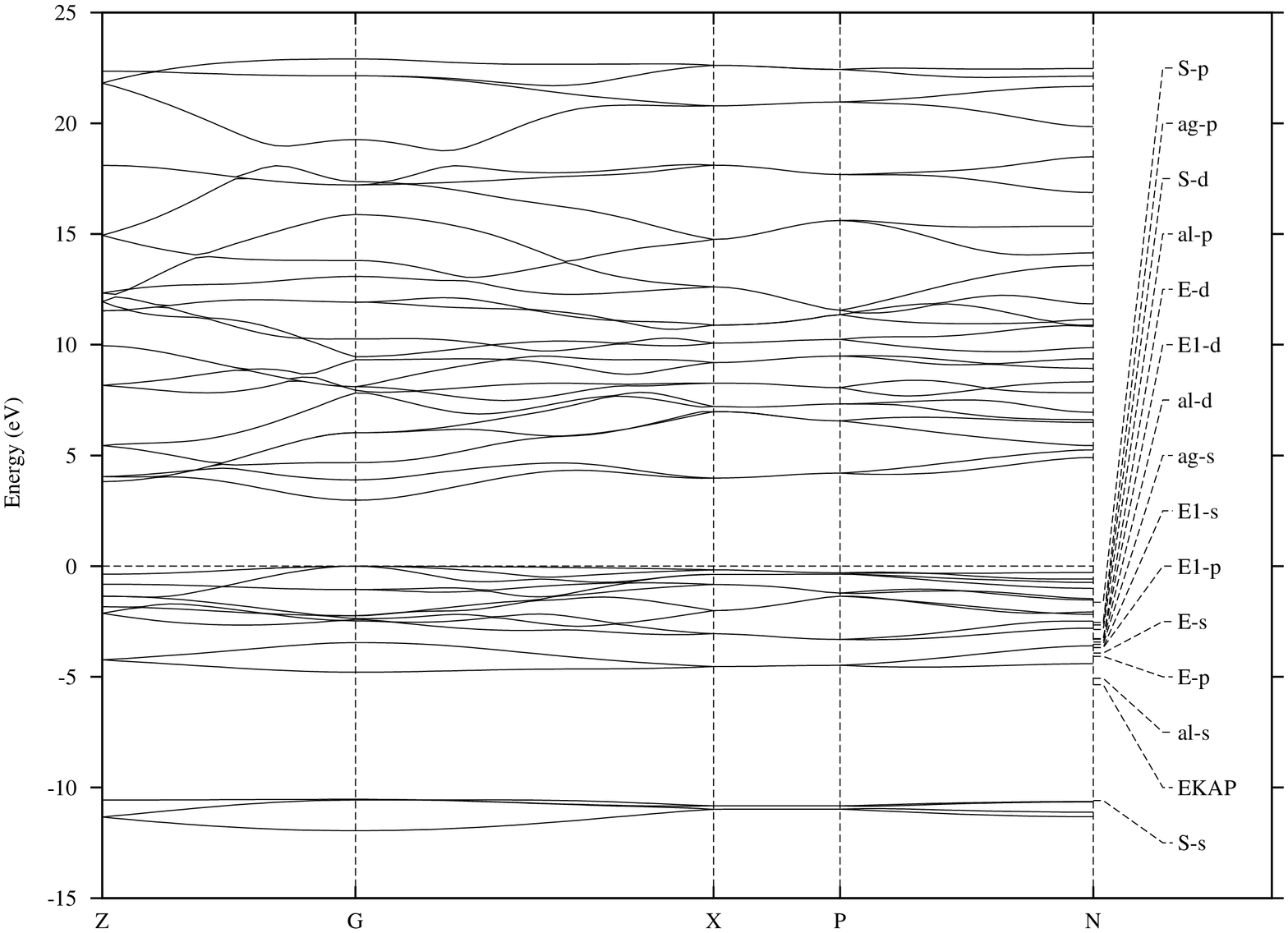}}} &
        \raisebox{1.3mm}[0pt]{\rotatebox{-90}{\resizebox{61.9mm}{!}{{\includegraphics{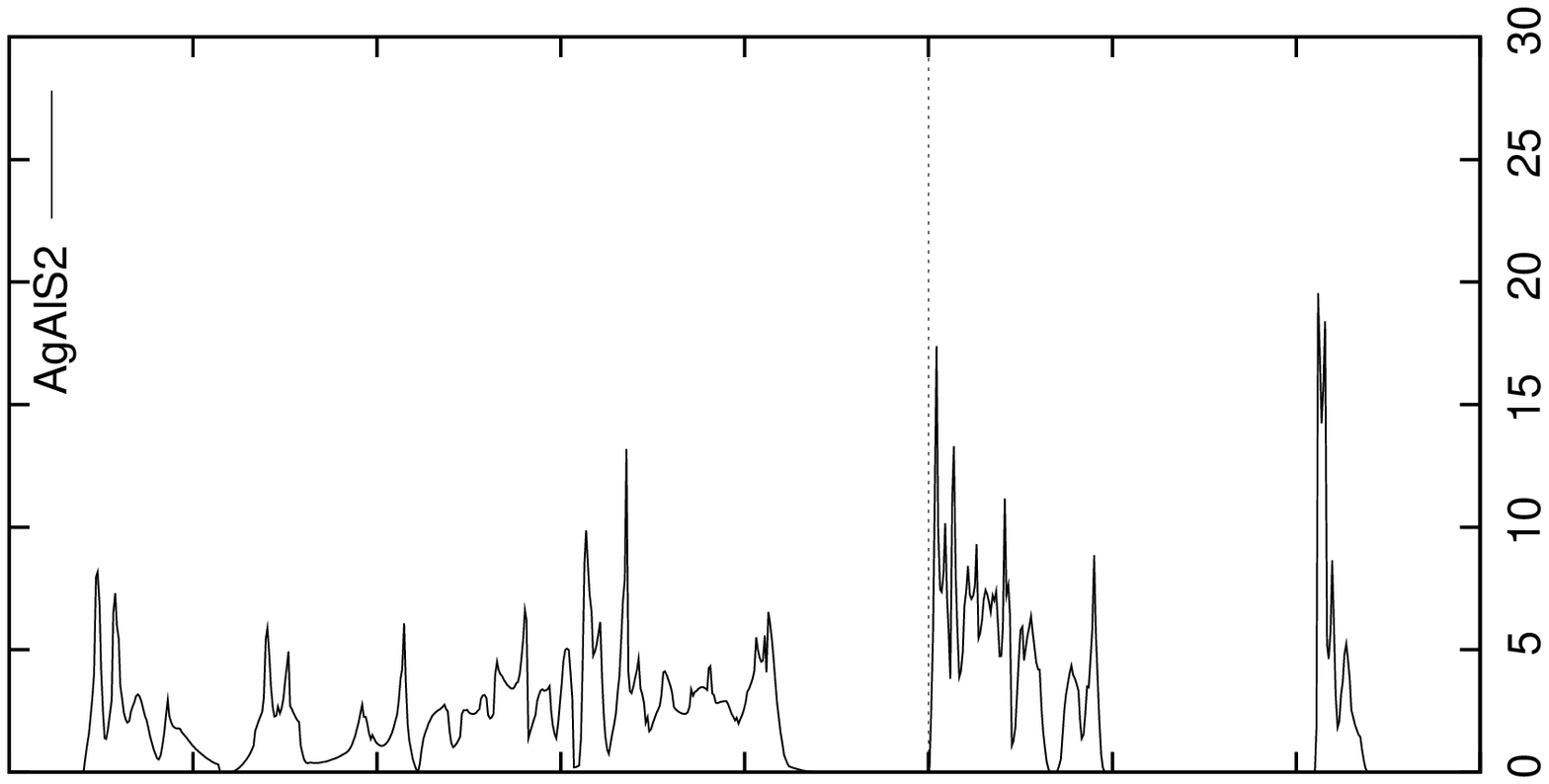}}}}} \\
        \end{tabular}
    \caption{Band structure and TDOS for ideal and without hybridization AgAlS2.}
    \label{test4}
  \end{center}
\end{figure}   
\begin{figure}
  \begin{center}
  \setlength{\tabcolsep}{-1.70cm}
    \begin{tabular}{cc}
      \rotatebox{-90} {\resizebox{60mm}{!}{\includegraphics{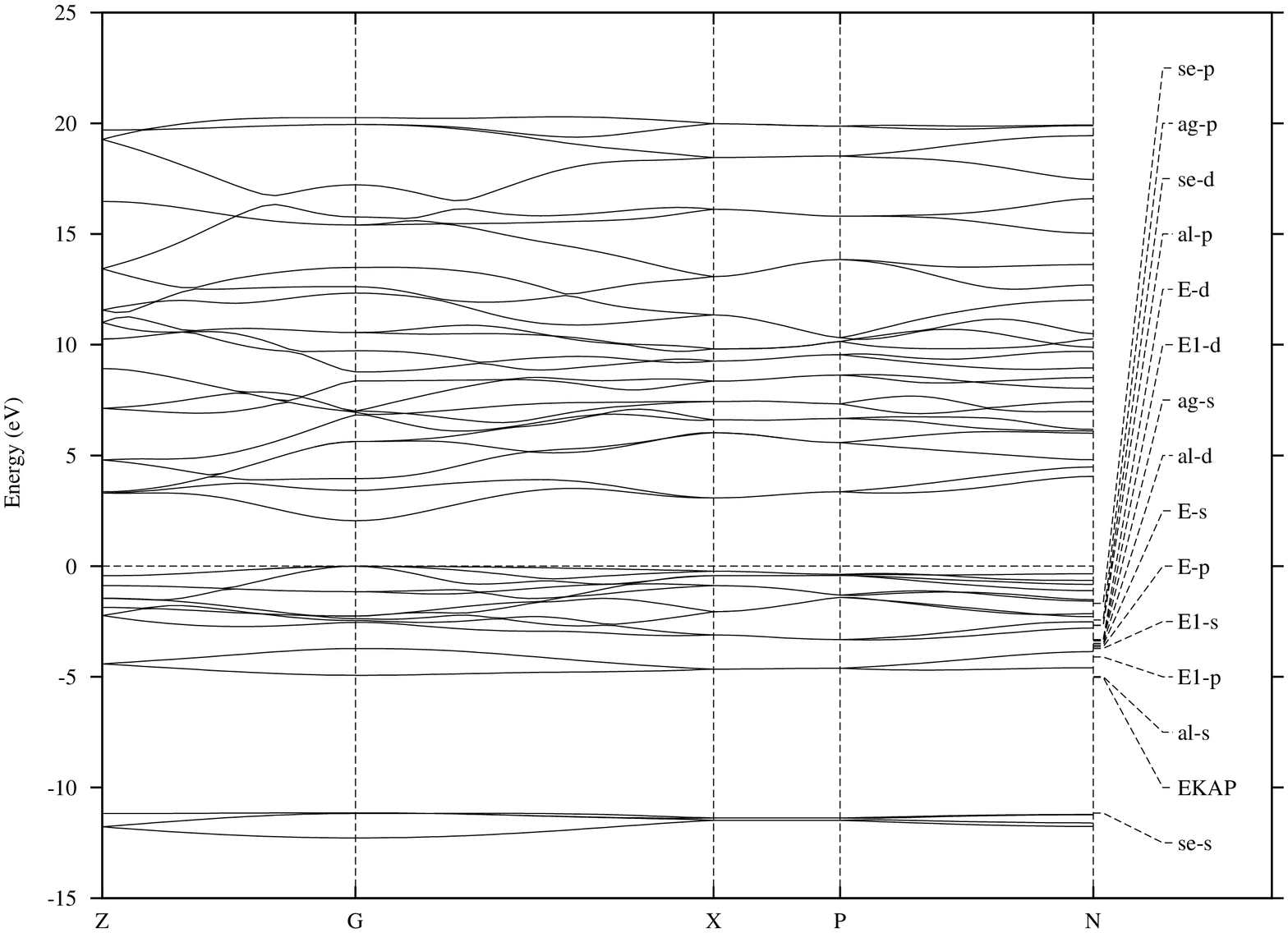}}} &
        \raisebox{1.3mm}[0pt]{\rotatebox{-90}{\resizebox{61.9mm}{!}{{\includegraphics{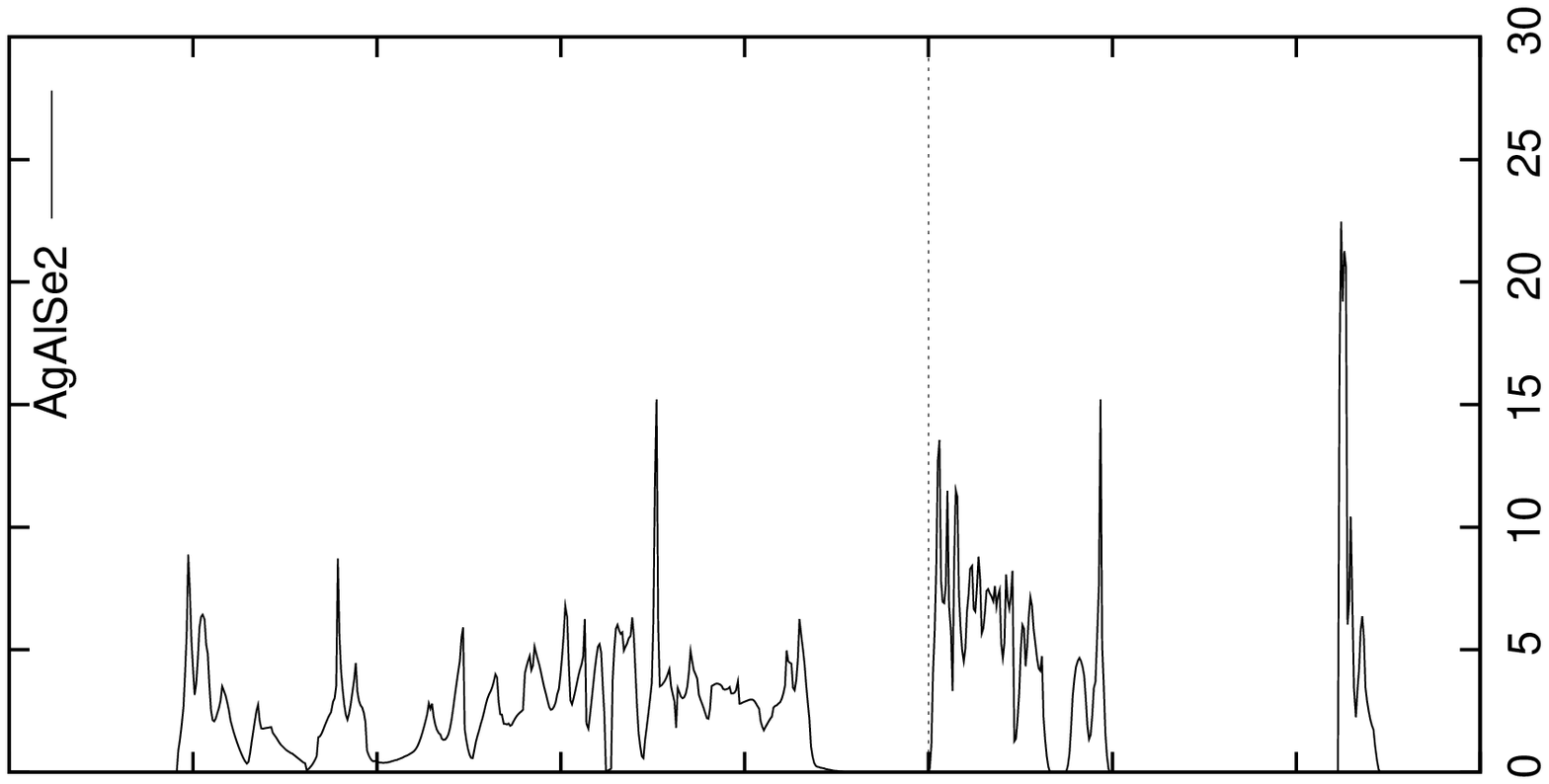}}}}} \\
        \end{tabular}
    \caption{Band structure and TDOS for ideal and without hybridization $AgAlSe_2$.}
    \label{test4}
  \end{center}
\end{figure}   
Unlike binary semiconductors like GaAs etc, the ternary semiconductor's `A' atoms in $A^IB^{III}C_2^{VI}$  belong to group I and are transition metals (Cu, Ag).
 \begin{figure}
  \begin{center}
  \setlength{\tabcolsep}{-1.70cm}
    \begin{tabular}{cc}
      \rotatebox{-90} {\resizebox{60mm}{!}{\includegraphics{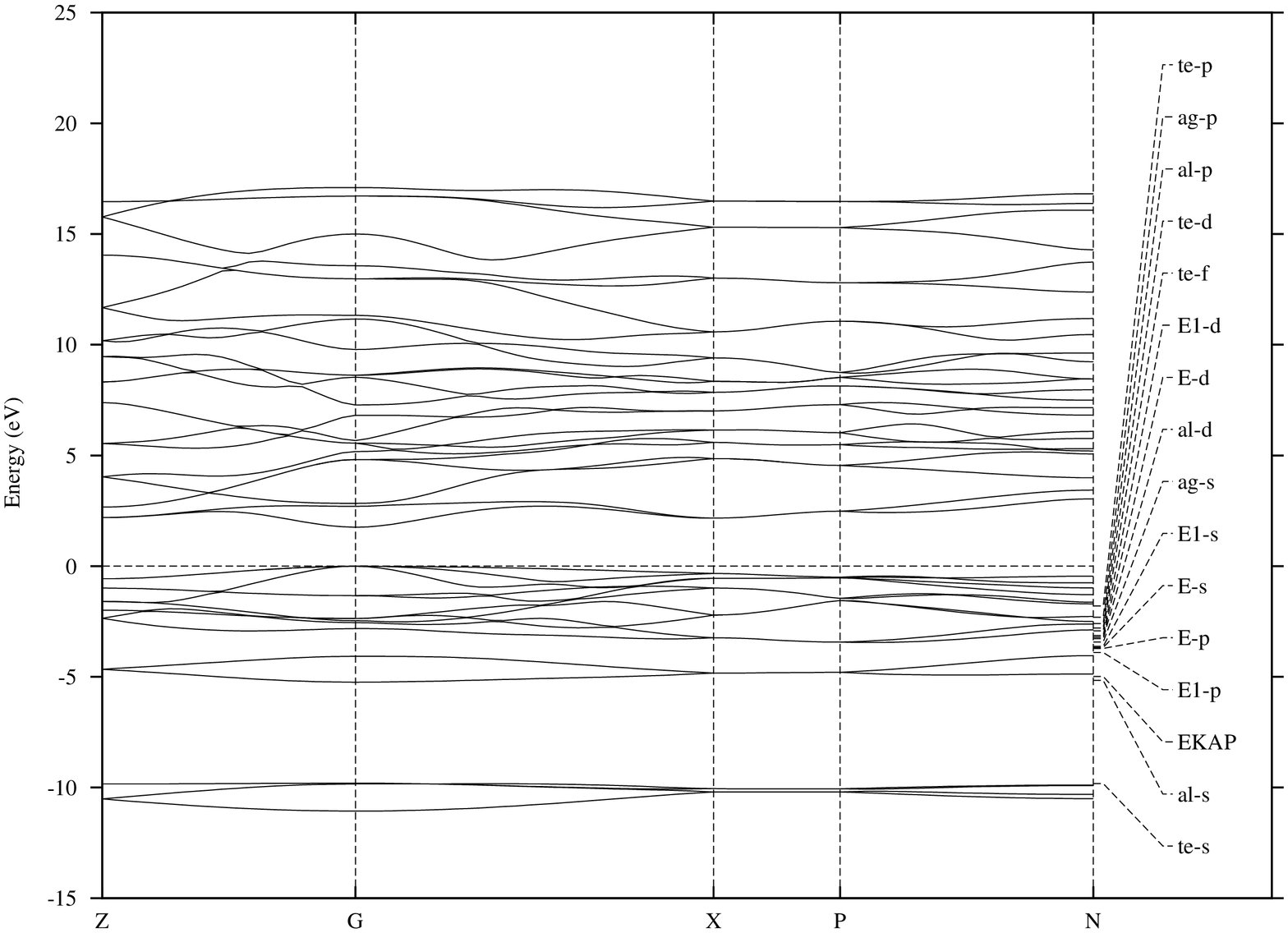}}} &
        \raisebox{1.3mm}[0pt]{\rotatebox{-90}{\resizebox{61.9mm}{!}{{\includegraphics{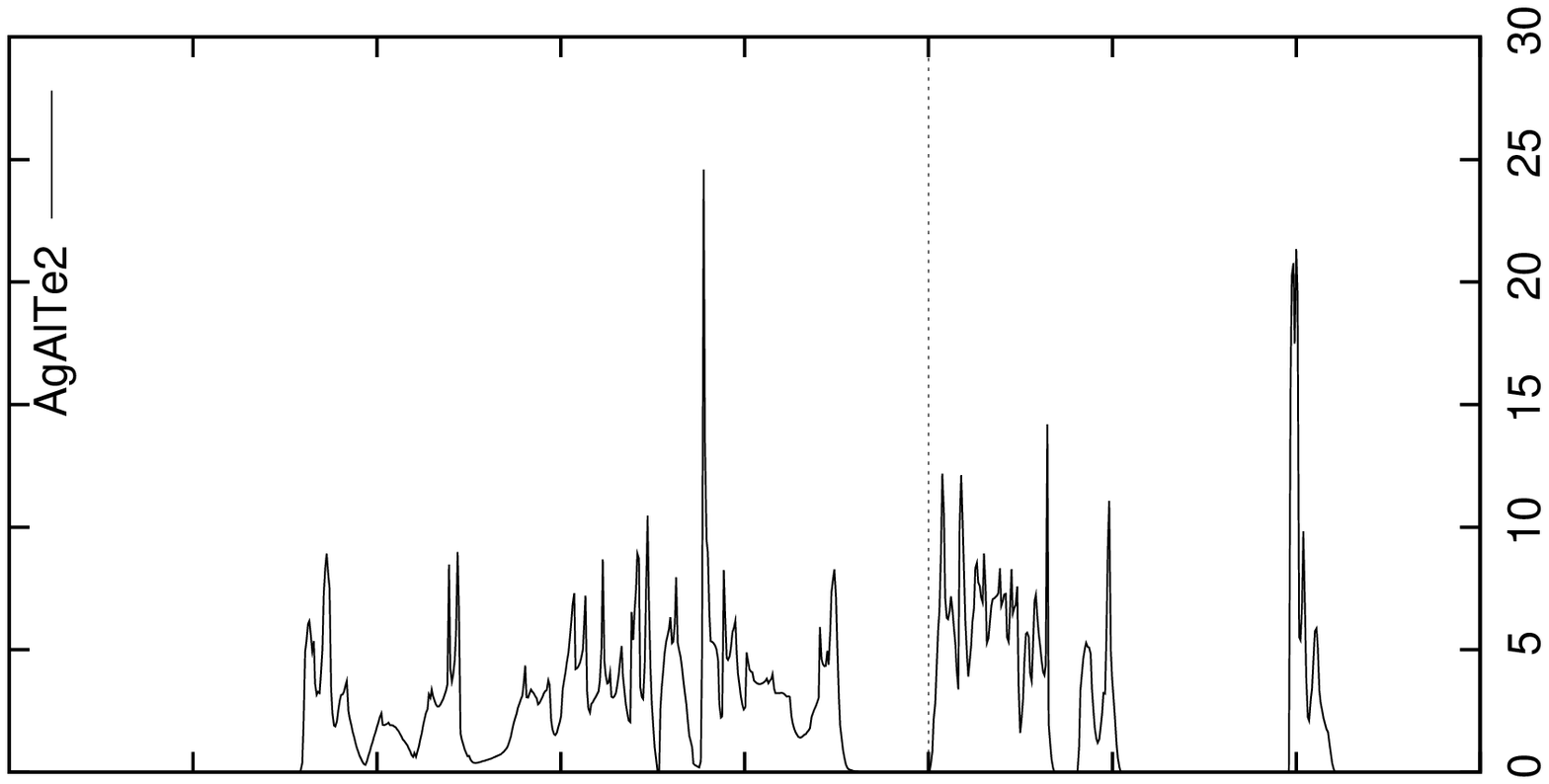}}}}} \\
        \end{tabular}
    \caption{Band structure and TDOS for ideal and without hybridization $AgAlTe_2$.}
    \label{test4}
  \end{center}
\end{figure}   
So these transition metal's d-orbitals also participate in forming valance bands. Therefore we expect Ag-d \& anion-p hybridization in the band formation in ternary semiconductors whereas there is strong s-p hybridization in case of binary semiconductors. 
It is observed in the case of Ag-based ternary compounds that p-d hybridization affect the band gap. In order to see explicitely the effect of p-d hybridization on the band gap and in general on electronic properties, we have calculated band structure and TDOS without the contribution of the d-orbitals for ideal $AgAlM_2$ systems. Therefor we have first freezed the d-electrons and have treated these electrons as core electrons. Figures 8,9,10 respectively show the band structure and TDOS with Ag-d electron as frozen for $AgAlM_2$. We have summarized the band gaps with and without contribution of d-electrons of Ag in table 2. The calculated result shows that there is a significant reduction of band gaps due to p-d hybridization and it is 51\% for $AgAlS_2$, 47\% for $AgAlSe_2$ and 42\% for $AgAlTe_2$.
\begin{table}[h]
\caption{\label{label} Reduction in band gap(eV) due to hybridization for ideal structure.}
\begin{tabular}{l@{}c@{}c@{}c@{}}\hline
 & \multicolumn{2}{|c|}{ Hybridization} &   \\ \cline{2-3} 
\raisebox{1.5ex}[0pt]{Systems}  & \vline\ With  & \vline\ Without &  \vline\ \raisebox{1.5ex}[0pt]{Reduction} \\ \hline
$AgAlS_{2}$ &  $1.70$ &  $3.46$ & $51\%$ \\
$AgAlSe_{2}$ & $1.42$ &  $2.67$ & $47\%$ \\
$AgAlTe_{2}$ &  $1.24$ &  $2.15$ & $42\%$ \\
\hline
\end{tabular}
\end{table} 
p-d hybridization not only has strong effect on band gap reduction, it also strongly affects the electronic structure in general as seen from the comparision of TDOS with (figure 11) and without (figures 8,9,10) p-d hybridisation for ideal case. The p-d hybridization can be interpreted on the basis of simple molecular orbital considerations \cite{15}. The p-orbitals that possess the $\Gamma_{15}$ symmetry hybridize with those of the d-orbitals that present the same sysmetry. This hybridization forms a lower bonding state and an upper antibonding state. The antibonding state that constitutes the top of the valence band is predominantly formed by higher energy anion p-states. And bonding state is constituted by the lower energy cation d-states. Perturbation theory \cite{25} suggests that the two states $\Gamma_{15}$(p) and $\Gamma_{15}$(d) will repel each other by an amount inversely proportional to the energy difference between p and d states. So this rising of the upper most state causes a gap reduction.
\subsection{Structural effect on electronic properties}
\label{3.4}
If we compare the TDOS of ideal $AgAlM_2$ (Figure 11) with that of non-ideal TDOS (figures 2,4,6), we observe that the structural distortion like bond alternation and tetragonal distortion are important factors which controls the band gap.
\begin{figure}[h]
\centering
\rotatebox{-90}{\resizebox{7.0cm}{10.0cm}{\includegraphics{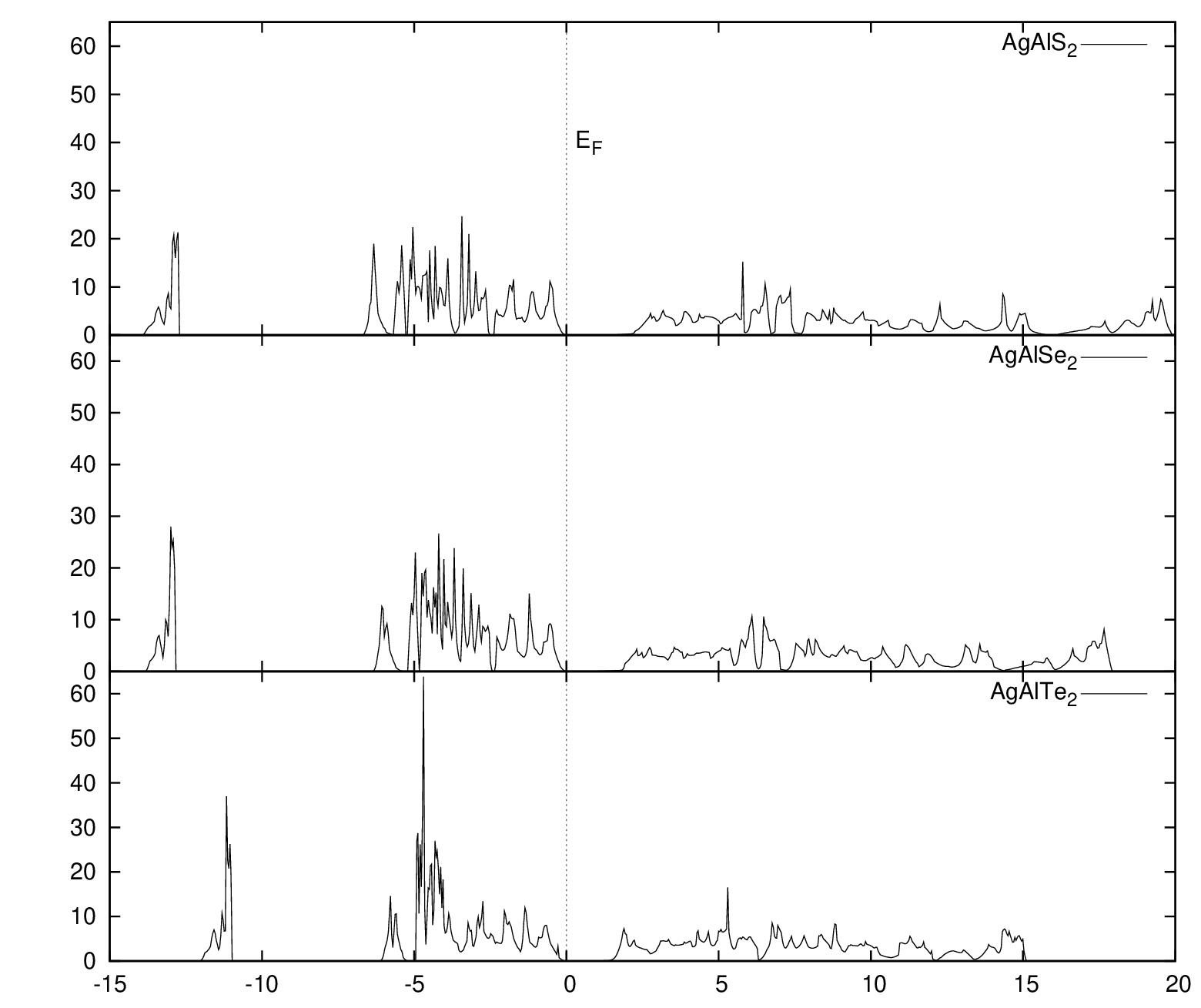}}}
\caption{TDOS for ideal AgAlS2, AgAlSe2 and AgAlTe2.}
\end{figure}
 
The structural contribution to the band gap is mainly controlled by bond alternation and has only a small contribution from tetragonal distortion. Similar effect on band gap is seen in $CuInSe_2$ \cite{15}. Important thing is to be noted that unlike in case of $CuInSe_2$ \cite{15} the bond alternation is responsible for the increase in band gap in the case of $AgAlM_2$. Table 3 shows the band gap increament in the three cases. The increament is maximum in the case of $AgAlS_2$ because anion displacement is maximum (u = 0.265) in this system. As u increases, the hybridization between Ag-d \& S-p decreases. Therefore the  occupied valence bands are stabilized by moving down in energy and the antibonding conduction bands are destabilized by moving up in energy. Thus band gap increases.\\
Our study shows that structural distortion not only affects the band gap but also the overall electronic structure significantly. A close comparision of TDOS for ideal (figure 11) and non-ideal (figures 2,4,6) show distinct differences in the structure in DOS. For example a very sharp peak is found at energy -3.79 eV for non-ideal $AgAlS_2$ (figure 2) compared to the corresponding ideal case (figure 11). This is due to individual character of Ag-d and S-p states. Figure 2 also shows that second and third bands get separated due to distortion in case of $AgAlS_2$ and hence increament of band gap by 0.11 eV. Similar result is observed in case of $AgAlSe_2$ and $AgAlTe_2$ as shown in figure 11. For example a sharp peak at $\simeq$  4.64 eV which appears in ideal $AgAlTe_2$ case disappears due to distortion. This is due to weaker p-d hybridization in non-ideal case. The width of the upper most valence band in case of $AgAlTe_2$ also decreases by 0.24 eV. There are effects on conduction band also. Like in the case of $AgAlS_2$ there is a sharp peak at 5.85 eV which disappears due to distortion. The conduction band width also decreases, 0.6 eV for $AgAlS_2$ and 0.36 eV for $AgAlTe_2$. The effect of distortion on valence and conduction bands shows that distortion is also responsible for significant change in optical properties of such semiconductors.
\begin{table}
\caption{\label{label} Increment in band gap(eV) due to structural distortion (without hybridization).}
\begin{tabular}{lccc}\hline
Systems &  Ideal  &  Non-ideal  &  Increment \\ \hline
$AgAlS_{2}$ &  $3.46$ &  $3.80$ & $9.8\%$ \\
$AgAlSe_{2}$ & $2.67$ &  $2.89$ & $8.2\%$ \\
$AgAlTe_{2}$ &  $2.15$ &  $2.26$ & $5.1\%$ \\
\hline
\end{tabular}
\end{table}
\section{Conclusion}
Calculations and study of $AgAlM_2$ (M = S, Se, Te) suggest that these compounds are direct band gap semiconductors with band gaps of 1.98 eV, 1.59 eV and 1.36 eV respectively. Our study further shows that electronic properties of these semiconductors significantly depend on structural distortion and the type of hybridization. The calculation is carried out using DFT based TB-LMTO method. we have used LDA for our exchange co-relation functional. Taking into account of the underestimation of band gap by LDA, our result of band gap and structutal properties agree with experimental values. Detail study of TDOS and PDOS shows that p-d hybridization between Ag-d and anion-p drastically reduces the band gap. The reduction is 51\%, 47\%, 42\% respectively for M = S, Se and Te. But band gap increases due to anion displacement in contrast to the result obtained by Jaffe and Zunger \cite{15} in case of $CuInSe_2$. This is due to the relative value of anion displacement parameter `u'. In the case of $AgAlM_2$, u increases whereas for $CuInSe_2$ it decreases with respect to the ideal value u = 0.25 corresponding to binary zinc blende structure. The gap enhancement due to anion distortion is 9.8\%, 8.2\% and 5.1\% respectively for $AgAlM_2$ (M = S, Se and Te). TDOS and PDOS further show that there is significant effect on the electronic properties due to structural distortion and the presence of p-d hybridization.
\section*{Acknowledgement}
This work was supported by Department of Science and Technology, India, under the grant no.SR/S2/CMP-26/2007. We would like to thank Prof. O.K. Andersen, Max Planck Institute, Stuttgart, Germany, for kind permission to use the TB-LMTO code developed by his group.

\end{document}